\documentclass[aps,twocolumn,english,preprintnumbers,amsmath,amssymb,nofootinbib,superscriptaddress]{revtex4-2}

\usepackage{hyperref}
\usepackage{graphicx}
\usepackage{xcolor}
\usepackage{bbold}

\usepackage{amsmath}
\newcommand{\beq}{\begin{equation}}
\newcommand{\eeq}{\end{equation}}
\newcommand{\bqa}{\begin{eqnarray}}
\newcommand{\eqa}{\end{eqnarray}}

\def\sumint{\hbox{$\sum$}\!\!\!\!\!\!\!\int}
\def\square{\vcenter{\vbox{\hrule height.4pt
          \hbox{\vrule width.4pt height4pt
          \kern4pt\vrule width.3pt}\hrule height.4pt}}}

\newcommand{\symff}{\text{SYM}_{4,4}}
\newcommand{\symot}{\text{SYM}_{1,10}}
\def\symod{\text{SYM}_{1,{\cal D}}}

\begin{document}

\title{${\cal N}=4$ supersymmetric Yang-Mills thermodynamics from effective field theory}

\author{Jens O. Andersen}
\affiliation{Department of Physics, 
Norwegian University of Science and Technology, H{\o}gskoleringen 5,
N-7491 Trondheim, Norway}

\author{Qianqian Du}
\affiliation{Department of Physics, Guangxi Normal University, Guilin, 541004, China}
\affiliation{Guangxi Key Laboratory of Nuclear Physics and Technology, Guilin, 541004, China}
\affiliation{Institute of Particle Physics and Key Laboratory of Quark and Lepton Physics (MOS), Central China Normal University, Wuhan, 430079, China}

\author{Michael Strickland}

\author{Ubaid Tantary}
\affiliation{Department of Physics, Kent State University, Kent, OH 44242, United States}

\date{\today}

\begin{abstract}
The free energy density of ${\cal N}=4$ supersymmetric Yang-Mills theory in four space-time
dimensions is derived through second order in the 't Hooft coupling $\lambda$ 
at finite temperature using effective-field theory methods. The contributions to the free energy
density at this order come from the hard scale $T$ and the soft scale $\sqrt{\lambda} T$.
The effects of the scale $T$ are encoded in the coefficients of an effective three-dimensional
field theory that is obtained by dimensional reduction at finite temperature. The effects
of the electric scale $\sqrt{\lambda} T$ are taken into account by perturbative calculations in the effective theory.
\end{abstract}


\maketitle

\section{Introduction}

The thermodynamics of ${\cal N}=4$ supersymmetric Yang-Mills theory in four dimensions ($\symff$) is of great interest since, at finite-temperature, the weak-coupling limit of this theory has many similarities with quantum chromodynamics (QCD).  Unlike QCD, however, in $\symff$ it is possible to make use of the AdS/CFT correspondence \cite{Maldacena:1997re} between gravity in anti-de Sitter space (AdS) and the large-$N_c$ limit of conformal field theories (CFT) on the boundary of AdS to obtain results for $\symff$ thermodynamics in the strong coupling limit.  In the large-$N_c$ limit one has \cite{Gubser:1998nz}
\bqa\label{eq:strong}
\frac{\cal S}{{\cal S}_{\textrm{ideal}}} =\frac{3}{4}\bigg[1+\frac{15}{8}\zeta(3)\lambda^{-3/2} + {\cal O}(\lambda^{-3}) \bigg] \, ,
\label{eq:sclimit}
\eqa
where $\lambda = N_c g^2$ is the `t Hooft coupling, $\cal S$ is the entropy density, and ${\cal S}_{\rm ideal} = 2 d_A \pi^2 T^3/3$ is the corresponding non-interacting entropy density (Stefan-Boltzmann limit) with $d_A = N_c^2-1$ being the dimension of the adjoint representation.

In the opposite limit of weak-coupling, this ratio has recently been computed through ${\cal O}(\lambda^2)$ \cite{resumsuper}.  In Ref.~\cite{resumsuper} the authors used the Arnold and Zhai method \cite{arnold1,arnold2} to perform the soft resummations necessary, finding a finite result which possessed non-analytic terms proportional to $\lambda^{3/2}$ and $\lambda^2\log\lambda$.  In this paper we revisit this calculation making use of effective field theory (EFT) techniques to perform the soft resummations.  Like Ref.~\cite{resumsuper} we will compute the $\symff$ thermodynamic functions to ${\cal O}(\lambda^2)$.  This will serve as a check on the calculation performed in Ref.~\cite{resumsuper}.

An added benefit of using EFT methods is that one can more easily extend the calculations of the thermodynamic functions to higher order in the `t Hooft coupling.  EFT methods have been applied to the computation of the resummed perturbative thermodynamics of a variety of theories, including QCD through ${\cal O}(\lambda^{5/2})$ \cite{braatenqcd} and $\symff$ through ${\cal O}(\lambda^{3/2})$ in \cite{tytgat}.  Here we extend the $\symff$ EFT calculation to ${\cal O}(\lambda^2)$ and present a systematic framework for computing the hard and soft contributions to the thermodynamic functions in supersymmetric Yang-Mills (SUSY).  Our final results agree with the results obtained in Ref.~\cite{resumsuper} up to a small correction to one term contributing at ${\cal O}(\lambda^2)$.  The difference is due to an incorrect assignment of the dimension ($4-2\epsilon$ versus $4$) in one of the soft contributions included in Ref.~\cite{resumsuper}.  We find that, taking into account this correction, both results agree exactly and that the corrected result is numerically very close to the result reported originally in Ref.~\cite{resumsuper}.\footnote{An erratum to \cite{resumsuper} has been submitted to account for this correction.}

To perform the calculation we make use of two types of dimensional reduction: (1) the equivalence between ten-dimensional $\symot$ and four-dimensional $\symff$ upon dimensional reduction, and (2) the additional dimensional reduction of $\symff$ to three dimensions that occurs at high temperatures.  We will refer to the first type of dimensional reduction as SUSY dimensional reduction and the second as high-temperature dimensional reduction.  In the case of the high-temperature dimensional reduction, one obtains a three-dimensional EFT that can be written in terms of dimensionally-reduced fields.  The construction of this high-temperature EFT preserves supersymmetry~\cite{supersimon}.
However, it is important to note that in supersymmetric theories, one must take some care with the dimensionality of the vector and spinor spaces describing the fields to ensure that supersymmetry is preserved by the regularization scheme used to evaluate Feynman diagrams.  For this purpose, we make use of the regularization by dimensional reduction (RDR) scheme \cite{Brink:1976bc,Gliozzi:1976qd,Siegel:1979wq,Siegel:1980qs,Capper:1979ns,Avdeev:1982xy}.  In this scheme, calculations proceed as in canonical dimensional regularization, except that the size of the representations of the bosonic and fermionic degrees of freedom are constrained to be integer valued.

Our paper is organized as follows. 
In Sec.~\ref{sec:susy} we introduce SUSY and dimensional reduction of $\symot$ to $\symff$.
In Sec.~\ref{sec:dimreduction}, we briefly discuss high-temperature dimensional reduction and the
effective field theory approach to finite-temperature field theory at weak coupling.
In Sec.~\ref{sec:params}, the parameters of the effective three-dimensional theory are determined and in 
Sec.~\ref{sec:eft}, the calculation of the soft part of free energy density is calculated using the
effective theory. In Sec.~\ref{sec:conclusions} we summarize. The relevant sum-integrals in four dimensions and 
integrals in three dimensions are listed in Appendix \ref{app:sumints} and \ref{app:3dints} for completeness.
The generalized Pad\'{e} approximant which interpolates between the known weak- and strong-coupling limits for large $N_c$ is listed in Appendix \ref{app:pade}.

{\em Notation:} In the full theory, we use lower-case letters for Minkowski space four-vectors, e.g., $p$, and upper-case letters for Euclidean space four-vectors, e.g., $P$.  In the dimensionally reduced EFT one has $p_0 = P_0 = 0$ and $p$ coincides with $|{\bf p}|$.  We use the mostly minus convention for the metric.

\section{Supersymmetric Yang-Mills theory}
\label{sec:susy}

We start with the action of $\mathcal{N}=1$ supersymmetric Yang-Mills in ${\cal D}$ dimensions  in Minkowski space~\cite{Brink:1976bc,Vazquez-Mozo:1999yck}
\beq \label{actN1}
S_{\textrm{SYM}_{1,{\cal D}}} = \int d^\mathcal{D} x \,  \textrm{Tr}\bigg[{-}\frac{1}{2}G_{\mu\nu}G^{\mu\nu}+2i\bar{\psi} \Gamma^\mu D_\mu \psi \bigg] \, ,
\eeq
where $\mu, \nu =0, \cdots, \mathcal{D}-1 $, and the field strength tensor is $G_{\mu\nu}=\partial_\mu A_\nu-\partial_\nu A_\mu-ig[A_\mu,A_\nu]$, and $D_\mu=\partial_\mu-i g[A_\mu,\cdot]$ is the covariant derivative in the adjoint representation of $SU(N_c)$. One can obtain supersymmetric field theories with  different number of supercharges, $n_{\rm SC}$,  by taking values of $\mathcal{D}$ for which the number of supercharges is maximal, resulting in
\bqa
n_{\rm SC}&=& 16 \rightarrow \mathcal{D}_{\textrm{max}} =10\;, \nonumber \\ 
n_{\rm SC} &=& 8 \rightarrow \mathcal{D}_{\textrm{max}} =6\;, \nonumber \\  
n_{\rm SC} &=& 4 \rightarrow \mathcal{D}_{\textrm{max}} =4\;.
\label{eq:nsc}
\eqa
To preserve supersymmetry, the number of bosonic and fermionic degrees of freedom must be equal.  One, 
therefore, needs to impose certain conditions on the fermions.   Thus, for $ \mathcal{D}_{\textrm{max}}=10$ fermions are Majorana-Weyl type, while for $ \mathcal{D}_{\textrm{max}}=6$ and $ \mathcal{D}_{\textrm{max}}=4$ they satisfy Weyl conditions. These constraints ensure that the number of bosonic and fermionic degrees of freedom are equal to $\mathcal{D}_{\textrm{max}}-2$.
We are in general interested in supersymmetric field theories with $n_{SC}$ supercharges in dimensions $D\leq \mathcal{D}_{\textrm{max}}$, with $D$ being an integer.
The evaluation of Feynman diagrams for theories that are obtained by SUSY dimensional reduction can be carried out in a simple way, in which we take all fields in Eq.~\eqref{actN1} to be  $\mathcal{D}$-dimensional tensors or spinors and all momentum to be $d=D-2\epsilon$ vectors~\cite{Vazquez-Mozo:1999yck}.

The $\symff$ theory can be obtained by dimensional reduction of $\symod$ in \mbox{$\mathcal{D}=\mathcal{D}_{\textrm{max}}=10$} with all fields being in the adjoint representation of $SU(N_c)$. The Minkowski space Lagrangian for $\symff$ can be expressed as \cite{Yamada:2006rx,Czajka:2012gq}
\bqa\label{lag}
\mathcal{L}_{\symff}&=& \textrm{Tr}
\bigg[{-}\frac{1}{2}G_{\mu\nu}G^{\mu\nu}
+(D_\mu\Phi_A)(D^\mu\Phi_A)
\nonumber\\
&& \hspace{1cm} +i\bar{\psi}_i {\displaystyle{\not} D}\psi_i  -\frac{1}{2}g^2(i[\Phi_A,\Phi_B])^2 
\nonumber\\
&& \hspace{1cm} - i g \bar{\psi}_i\big[\alpha_{ij}^{\texttt{p}} X_{\texttt{p}}+i \beta_{ij}^{\texttt{q}}\gamma_5Y_{\texttt{q}},\psi_j\big] \bigg] 
\nonumber\\
&& \hspace{1cm} + \mathcal{L}_{\textrm{gf}}+\mathcal{L}_{\textrm{gh}}+\Delta\mathcal{L}_{\symff} \, ,
\eqa
where $\psi_i$ represents four Majorana spinors, $G_{\mu\nu}^a=\partial_\mu A_\nu^a-\partial_\nu A_\mu^a+gf^{abc}A_\mu^bA_\nu^c$ is the nonabelian field strength with gauge coupling $g$, and $\mu$, $\nu=0,1,2,3$.  There are six scalar fields with $\Phi_A=(X_1,Y_1,X_2,Y_2,X_3,Y_3)$, where $X_{\texttt{p}}$ and $Y_{\texttt{q}}$ are Hermitian matrices, with ${\texttt{p,q}}=1,2,3$. $X_{\texttt{p}}$ and $Y_{\texttt{q}}$ represent scalar and pseudoscalar fields, respectively.  The $4\times 4$ matrices $\alpha^{\texttt{p}}$ and $\beta^{\texttt{q}}$ satisfy
\beq\label{lag1}
\{\alpha^{\texttt{p}},\alpha^{\texttt{q}}\}=-2\delta^{\texttt{pq}} \, , \;\; \{\beta^{\texttt{p}},\beta^{\texttt{q}} \}=-2\delta^{\texttt{pq}} \, , \;\; [\alpha^{\texttt{p}},\beta^{\texttt{q}}]=0 \, .
\eeq
An explicit representation of these matrices can be found in Sec.~3 of Ref.~\cite{resumsuper}.

The ghost term $\mathcal{L}_{\textrm{gh}}$ depends on the choice of the gauge-fixing term $\mathcal{L}_{\textrm{gf}}$.  Here we work in general covariant gauge, giving
\bqa\label{gf}
\mathcal{L}_{\textrm{gf}}^{\symff} &=& -\frac{1}{\xi}\textrm{Tr}\big[(\partial^\mu A_\mu)^2 \big],   \\
\mathcal{L}_{\textrm{gh}}^{\symff} &=&-2\textrm{Tr}\big[\bar{\eta}\,\partial^\mu \! D_\mu \eta \big] ,
\label{gh}
\eqa
with $\xi$ being the gauge parameter.  Finally, the last term in Eq.~\eqref{lag}, $\Delta\mathcal{L}_{\symff}$, represents any counterterms necessary to renormalize the theory.

\section{Dimensional reduction at finite temperature}
\label{sec:dimreduction}

Dimensional reduction and the effective-field theory approach to field theory at high
temperature is now well 
established~\cite{ginsparg,landsman,farakos1,farakos2,farakos3,finnseft,braatenscalar,braatenqcd,york}.
The idea is as follows. In the imaginary-time formalism, loop diagrams involve summations of
Matsubara frequencies and integrals over three-momenta. These frequencies are $2n\pi T$ for bosons and $(2n+1)\pi T$ for fermions, where $n$ is an integer. These
frequencies act as masses in the propagators and thermal field theory in equilibrium can be
considered a Euclidean three-dimensional field theory with an infinite tower of massive modes, except the zeroth ($n=0$) bosonic mode, which is massless. 

Screening effects in the medium generate
a thermal mass (static electric screening) of the order 
$gT$, where $g$ is a generic coupling. At weak coupling, the momentum scales
$T$ and $gT$ are well separated and one expects the nonzero Matsubara modes to decouple at high
temperature.\footnote{High temperature means we can ignore any zero-temperature masses in the theory.}
The static correlators of the full theory can be reproduced to any desired accuracy at length scales
$R\gg1/T$ by matching an effective three-dimensional theory for the zero modes to the full theory by fixing the parameters in the EFT to be functions of the temperature and the parameters of the full theory.
The couplings encode the physics on the scale $T$, while contributions to physical quantities
on the scale $gT$ are taken care of by calculations in the low-energy effective theory.
In nonabelian gauge theories, there is a third, supersoft scale on the order of $g^2T$ associated
with screening of static (chromo)magnetic fields. It may therefore prove useful to integrate out
the masses of order $gT$ to construct a second effective field theory valid on the momentum scale  $g^2T$. However, perturbation theory is
plagued with infrared divergences since static magnetic fields are not screened
and nonperturbative methods such as lattice simulations must be 
used~\cite{linde,gross}.
In the present case, as in QCD~\cite{braatenqcd}, this construction is only necessary if one is interested in the free energy density at order $\lambda^3$.

The starting point is the partition function given as a path integral in the full theory,
\beq
{\cal Z}_{\symff}= \! \int \!
{\cal D}\bar{\eta}{\cal D}\eta
{\cal D}\bar{\psi}_i{\cal D}\psi_i
{\cal D}A_{\mu}^a{\cal D}\Phi_A^a \, e^{- \!{\int_0^{\beta}}\! d\tau\int d^3x \, {\cal L}_{\symff}}\,,
\eeq
where $\bar{\eta}$, $\eta$, $\bar{\psi}_i$, $\psi_i$, $A_{\mu}^a$, and $\Phi_A^a$ are the fields
in the Lagrangian Eqs.~(\ref{lag})--(\ref{gh}).

The free energy is then given by
\bqa
{\cal F}&=&-{1\over\beta V}\log{\cal Z}_{\symff}\;,
\label{zsym}
\eqa
where $V$ is the volume of space. 
After having integrated out the non-static modes, we can write
the partition function as
\beq
{\cal Z} = \int{\cal D}\bar{\eta}{\cal D}{\eta}{\cal D}A_0^a{\cal D}A_i^a{\cal D}\Phi_A^ae^{-f_EV-\int \!d^3x\,{\cal L}_{\rm ESYM}}\;,
\label{zesym}
\eeq
where $\bar{\eta}$, $\eta$,
$A_0^a$, $A_i^a$, and $\Phi_A^a$ are fields in the effective theory. 
Up to normalizations,
the fields in the effective theory can be identified with the fields in the original theory.
$f_E$ is the coefficient of the unit operator and can be interpreted as the
contribution to the free energy from the hard scale $T$. 
${\cal L}_{\rm ESYM}$ is given by the most general Lagrangian that can be constructed from the
fields $A_i^a$, $A_0^a$, and $\Phi_A^a$. We find
\bqa\nonumber
{\cal L}_{\rm ESYM}&=&
\frac{1}{2}{\rm Tr}\left[G_{ij}^2\right]
+{\rm Tr}[(D_iA_0)(D_iA_0)]
\nonumber \\ && 
+{\rm Tr}[(D_i\Phi_A)(D_i\Phi_A)]
+m_E^2{\rm Tr}[A_0^2]
\nonumber \\ &&
+m_S^2{\rm Tr}[\Phi_A^2]
+h_E{\rm Tr}[(i[A_0,\Phi_A])^2]
\nonumber \\ && 
+\frac{1}{2}g^2_3{\rm Tr}[(i[\Phi_A,\Phi_B])^2]
+{\cal L}_{\rm gf}+{\cal L}_{\rm gh}+
\delta{\cal L}_{\rm ESYM} \, , \nonumber \\
\label{lagesym}
\eqa
where $A_0=t^aA_0^a$, $\Phi_A=t^a\Phi_A^a$, $(D_iA_0)^a=\partial_iA_0^a+g_E f^{abc}A_i^b A_0^c$, and
$G_{ij}^a=\partial_i A_j^a-\partial_j A_i^a+g_Ef^{abc}A_i^bA_j^c$ is the nonabelian
field strength with gauge coupling $g_E$ and $f^{abc}$ are the structure constants.  We work in general covariant gauge, where the gauge-fixing and ghost terms
are given by
\bqa
{\cal L}_{\rm gf}&=&-{1\over\xi}{\rm Tr}\left[(\partial_{i}A_{i})^2\right]
,\\
{\cal L}_{\rm gh}&=&-2{\rm Tr}[\bar{\eta}\partial_{i}D_{i}\eta]
\;,
\eqa
where $\eta$ is the ghost field and $\xi$ is the gauge parameter.
Finally, ${\delta}{\cal L}_{\rm ESYM}$ represents all higher-order local operators that can be constructed
from $A_{i}^a$, $A_0^a$, and $\Phi_A^a$
satisfying the symmetries, such as gauge invariance and rotational
invariance. For example, there will be two quartic self-interaction terms of $A_0$, namely
$({\rm Tr}[A_0^2])^2$ and ${\rm Tr}[A_0^4]$, however, they first contribute at order $\lambda^3$
to the free energy density.
The Lagrangian (\ref{lagesym}) is the same as that of Ref.~\cite{tytgat}, except
that we have explicitly shown the quartic self-coupling of the scalar fields $\Phi_A^a$
and their couplings to the adjoint field $A_0^a$ because these terms will generate
contributions to the free energy of order $\lambda^2$. 

\begin{figure*}[tbh]
\begin{center}
\includegraphics[width=0.725\linewidth]{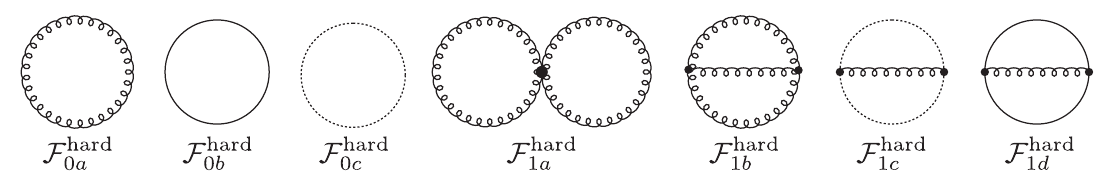}
\end{center}
\vspace{-4mm}
\caption{One- and two-loop diagrams contributing to the  ${\rm \symot}$ free energy density. Together with SUSY dimensional reduction from 10 to 4 dimensions these can be used to compute the one- and two-loop hard contributions in $\symff$.  Spiral lines represent ten-dimensional gluons, solid lines represent ten-dimensional Majorana-Weyl fermions, and dotted lines represent ten-dimensional ghost fields.}
\label{fig:diagrams12-10d}
\end{figure*}

\section{Parameters of the effective theory}
\label{sec:params}

In this section, we determine the parameters of the effective theory to the order
in the coupling $g^2$ which is needed to calculate the free energy to order
$\lambda^{2}$. In the matching calculations, we will be using so-called 
{\it strict perturbation theory}~\cite{braatenscalar,braatenqcd}.
In strict perturbation theory, we also treat the quadratic mass terms in the effective Lagrangian
as a perturbation. We also do not add and subtract a thermal mass term in the full theory to screen infrared divergences. In other words, we are using massless propagators in both the
full and effective theory in the perturbative calculations.
Such calculations are plagued by infrared divergences, but they appear in the same way on both sides of the matching equations and hence they cancel. Although this is an incorrect treatment of the
infrared divergences, we can use dimensional regularization to regulate them in this intermediate step of the calculations. The point is that the coefficients of the effective theory encode the physics on the scale $T$ and that a correct treatment of IR divergences is ensured by using massive propagators when we do perturbative calculations in the effective theory.

\subsection{Coefficient of the unit operator}

Equating Eqs.~(\ref{zsym}) and~(\ref{zesym}), and taking the logarithm, we obtain
\bqa
f_EV-{\log{\cal Z}_{\rm ESYM}}=-{\log{\cal Z}_{\symff}}\;.
\eqa
The right-hand side of this equation is given in terms of the vacuum diagrams of the full theory
using massless (unresummed) propagators. These are listed below through three loops.
The left-hand side is given by the coefficient of the unit operator and the vacuum diagrams
in the effective theory. Since we are using massless propagators in strict perturbation theory,
there is no scale in the momentum integrals and they are therefore set to zero in dimensional regularization. This implies $f_E V\approx-{\log{\cal Z}_{\rm SYM}}$, where $\approx$
is a reminder that the right-hand-side is obtained in strict perturbation theory.
We can therefore interpret $Tf_E$ as the unresummed or hard
contribution to the free energy density.

The diagrams in the full theory through three loops were evaluated in Ref.~\cite{resumsuper}. For the one- and two-loop graphs, the authors of Ref.~\cite{resumsuper} calculated directly in $\symff$ because the thermal mass contributions had to be computed and it was not possible to use SUSY dimensional reduction from ${\rm \symot}$ for this purpose. In this paper, for the unresummed (hard) contributions we do not need to consider the thermal masses of the gluons, fermions, or scalars.  As a result, using the EFT method, we can calculate the hard contributions using SUSY dimensional reduction from ${\rm \symot}$ to $\symff$.  This allows us to compute a reduced number of $\symod$ diagrams for general ${\cal D}$ and $d$, from which we can obtain the $\symff$ result by taking $ {\cal D} = 10 $, $D=4$, and the number of momentum-space dimensions to be $ d = 4-2 \epsilon $~\cite{Vazquez-Mozo:1999yck}.

We list all the three loop results here for completeness.  The one- and two-loop graphs in $\symod$ are shown in Fig.~\ref{fig:diagrams12-10d}.  Summing the one-loop graphs ${\cal F}^{\rm hard}_{0a}$,  ${\cal F}^{\rm hard}_{0b}$, and  ${\cal F}^{\rm hard}_{0c}$, one obtains 
\bqa
{\cal F}_{0}^{\rm hard}&=&
\frac{1}{2} d_A ({\cal D}-2) ( f_0^{\prime} -b_0^{\prime}) 
=-d_A{\pi^2\over6}T^4\;,
\label{f11}
\eqa
where $b_0^{\prime}$ and $f_0^{\prime}$ are defined in Eqs.~(\ref{b0}) and~(\ref{f0}).  

Summing the two-loop graphs shown in Fig.~\ref{fig:diagrams12-10d}, one obtains
\bqa
\nonumber
{\cal F}_1^{\rm hard}&=&d_A\lambda\left[{\cal F}_{1a}^{\rm hard}+
{\cal F}_{1b}^{\rm hard}+{\cal F}_{1c}^{\rm hard}+{\cal F}_{1d}^{\rm hard} \right] ,
\eqa
with the individual contributions being
\bqa
{\cal F}_{1a}^{\rm hard}&=&{{\cal D} ({\cal D}-1)\over4}b_1^2\;,\\
{\cal F}_{1b}^{\rm hard}&=&-{3\over4}({\cal D}-1)b_1^2\;,\\
{\cal F}_{1c}^{\rm hard}&=& {1\over4}b_1^2\;,\\
{\cal F}_{1d}^{\rm hard}&=& {({\cal D}-2)\over4} {\rm Tr}\,\mathbb{1} \left[ f_1^2-2f_1 b_1 \right]\;.
\eqa
Above ${\rm Tr}\,\mathbb{1}$ is the dimension of the spinors in the maximal SYM theory which equals ${\rm Tr}\,\mathbb{1} = {\cal D}-2$ for all cases listed in Eq.~\eqref{eq:nsc}. The integrals $b_1$ and $f_1$ are defined in Eqs.~(\ref{b1}) and~(\ref{f1}).
This yields
\bqa\nonumber
{\cal F}_{1}^{\rm hard}&=&
d_A\lambda{({\cal D}-2)^2\over4}\left(b_1-f_1\right)^2
\\ &=&
\label{f22}
d_A\left({\pi^2T^4\over6}\right){3\over2}{\lambda\over\pi^2}\;.
\eqa

\begin{figure*}[t]
\begin{center}
\hspace{-1mm}\includegraphics[width=1.01\linewidth]{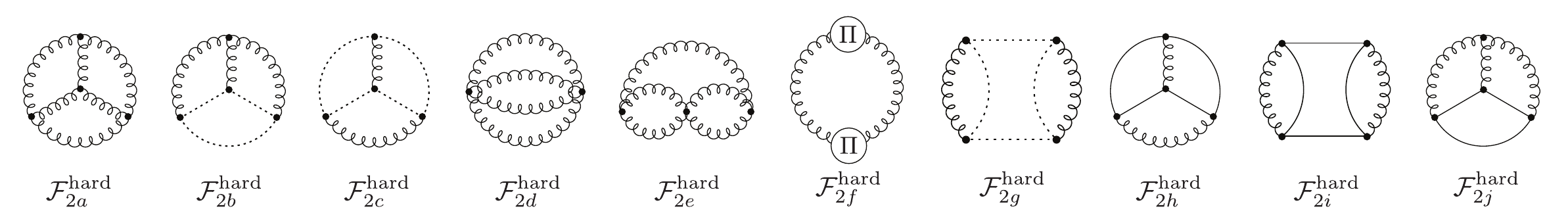}
\end{center}
\vspace{-5mm}
\caption{Three-loop diagrams contributing to the ${\rm \symot}$ free energy.  Notation is the same as Fig.~\ref{fig:diagrams12-10d}.  A circle with $\Pi$ in it represents the one-loop gauge field self-energy in ${\rm \symot}$.}
\label{fig:diagrams3}
\end{figure*}

The three-loop graphs in $\symod$ are shown in Fig.~\ref{fig:diagrams3} with
\bqa\nonumber
{\cal F}_2^{\rm hard}&=&d_A\lambda^2\left[{\cal F}_{2a}^{\rm hard}+
{\cal F}_{2b}^{\rm hard}+{\cal F}_{2c}^{\rm hard}+{\cal F}_{2d}^{\rm hard}
+{\cal F}_{2e}^{\rm hard}
\right.\\ &&\left.
+{\cal F}_{2f}^{\rm hard}+{\cal F}_{2g}^{\rm hard}
+{\cal F}_{2h}^{\rm hard}
+{\cal F}_{2i}^{\rm hard}+{\cal F}_{2j}^{\rm hard}\right].
\nonumber \\ &&
\eqa
The individual contributions were calculated in Ref.~\cite{resumsuper},
\bqa
{\cal F}_{2a}^{\rm hard}&=&\left[-{5{\cal D}\over8}+{23\over32}\right]I_{\rm ball}^{\rm bb}\;,
\\
{\cal F}_{2b}^{\rm hard}&=&
{1\over16}I_{\rm ball}^{\rm bb}\;,\\
{\cal F}_{2c}^{\rm hard}&=&
{1\over32}I_{\rm ball}^{\rm bb}\;,\\
{\cal F}_{2d}^{\rm hard}&=&
-{3\over16}{\cal D}({\cal D}-1)I_{\rm ball}^{\rm bb}\;,\\
{\cal F}_{2e}^{\rm hard}&=&
{27\over16}({\cal D}-1)I_{\rm ball}^{\rm bb}\;,
\\
{\cal F}_{2f}^{\rm hard}&=&-{1\over4}
\left[I_{\symod}^{\rm bb}+I_{\symod}^{\rm bf}
+I_{\symod}^{\rm ff}
\right] ,\\
{\cal F}_{2g}^{\rm hard}&=&
{1\over8}I_{\rm ball}^{\rm bb}\;,\\
{\cal F}_{2h}^{\rm hard}&=&
{{\cal D}-2\over8}{\rm Tr}\,\mathbb{1}\left[{{\cal D}-6\over2}I_{\rm ball}^{\rm ff}
+(4-{\cal D})I_{\rm ball}^{\rm bf}
\right],\\
{\cal F}_{2i}^{\rm hard}&=&
{({\cal D}-2)^2\over4}{\rm Tr}\,\mathbb{1}\left[I_{\rm ball}^{\rm bf}-2H_3+f_2(f_1-b_1)^2\right] , \nonumber \\ && \\
{\cal F}_{2j}^{\rm hard}&=&
-{{\cal D}-2\over4}{\rm Tr}\,\mathbb{1} \, I_{\rm ball}^{\rm bf} \; .
\eqa
Using the expressions for the sum-integrals listed in Appendix~\ref{app:sumints}, the three-loop hard contribution becomes
\bqa
{\cal F}_2^{\rm hard}&=&
-d_A\left({\pi^2T^4\over6}\right)\left[
{3\over4\epsilon}+{9\over2}\log{\Lambda\over4\pi T}
\right. \nonumber \\ && \label{f33} \left.
+{3\over2}\gamma_E
+3{\zeta^{\prime}(-1)\over\zeta(-1)}
+{15\over4}-\log2
\right]\left({\lambda\over\pi^2}\right)^2 \!. \;\;\;\;
\eqa
We note that there is a remaining pole in $\epsilon$ proportional to $\lambda^2$.
The pole is cancelled by the counterterm $\delta f_E$ for the coefficient of the unit operator $f_E$, which can be found by calculating the ultraviolet divergences in the effective theory~\cite{braatenqcd}.
Using dimensional regularization and minimal subtraction, the counterterm must be a polynomial
in $m_E$, $m_S$, $g_E$, $h_E$, $g_3$, and the other parameters of the ESYM.
The counterterm that cancels this divergence is
\bqa
\delta f_E&=&-{d_AN_c\over4(4\pi)^2\epsilon}g_E^2\left[m_E^2+6m_S^2\right],
\eqa
which is found by a two-loop calculation in the effective theory (see Sec.~\ref{sec:eft} below).
Since the mass parameters $m_E^2$ and $m_S^2$ multiply the pole in $\epsilon$, we must
take into account the order-$\epsilon$ contribution, when we express the counterterm in 
terms of the parameters $\lambda$ and $T$. Using the expressions for the mass parameters, Eqs.~(\ref{mpi1}) and~(\ref{ms1}) below,
the result is then
\begin{widetext}
\bqa\nonumber
\delta f_E&=&-{2d_A\lambda^2T\over(4\pi)^2\epsilon}(d+4)(b_1-f_1)
\\
&=&
\label{countf}
-d_A\left({\pi^2T^3\over6}\right)
\left[{3\over4\epsilon}+{3\over2}\log{\Lambda\over4\pi T}
+{21\over16}
+{3\over2}{\zeta^{\prime}(-1)\over\zeta(-1)}-{1\over2}\log2\right]\left({\lambda\over\pi^2}\right)^2\;.
\eqa
The final result for the renormalized unit operator $f_E$ is given by the
sum of Eqs.~(\ref{f11}),~(\ref{f22}),~(\ref{f33}), and~(\ref{countf}) 
\bqa\nonumber
f_E(\Lambda)T&=&
{\cal F}_0^{\rm hard}+{\cal F}_1^{\rm hard}+
{\cal F}_2^{\rm hard}-T\delta f_E
\\ &=&
\label{renhard}
-d_A{\pi^2T^4\over6}\left\{
1-{3\over2}{\lambda\over\pi^2}+\left[3\log{\Lambda\over4\pi T}
+{39\over16}+{3\over2}\gamma_E
+{3\over2}{\zeta^{\prime}(-1)\over\zeta(-1)}-{1\over2}\log2
\right]\left({\lambda\over\pi^2}\right)^2\right\} .
\eqa
\end{widetext}
The coupling $\lambda$ does not get  renormalized and is therefore independent of the
scale $\Lambda$, which implies that $f_E(\Lambda)$ is running.
Its running is given by the evolution equation
\beq
\Lambda{d\over d\Lambda}f_{E}(\Lambda)=
-{d_AN_c\over(4\pi)^2}g_E^2\left[m_E^2+6m_S^2\right],
\eeq
whose solution is 
\beq
f_{E}(\Lambda_{\rm })=f_{E}(\Lambda^{\prime})
-{d_AN_c\over(4\pi)^2}g_E^2\left[m_E^2+6m_S^2\right]
\log{\Lambda_{\rm }\over\Lambda^{\prime}_{\rm }}\;.
\eeq

\subsection{Mass parameters}

We need the mass parameters squared $m_E^2$ and $m_S^2$
for the adjoint field $A_0^a$ and the scalars $\Phi_A^a$ to one-loop order.
Their physical interpretation is that they give the contribution to the static screening masses
from the hard scale $T$. In non-abelian gauge theories and beyond leading order, the electric screening (or Debye) mass $m_E^2$ is plagued with infrared divergences associated with the lack of magnetostatic screening.
It therefore requires a nonperturbative definition~\cite{nonpert}. However, the hard contribution to the Debye mass can be computed order by order 
in strict perturbation theory.

\begin{figure}[tbh]
\begin{center}
\includegraphics[width=0.9\linewidth]{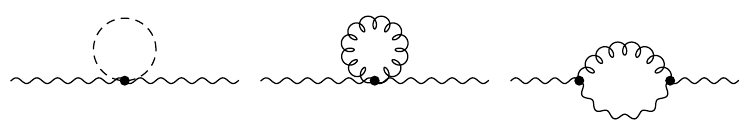}
\end{center}
\vspace{-5mm}
\caption{One-loop $A_0$ self-energy graphs in the $\symff$ dimensionally-reduced EFT.  
Spiral lines represent three-dimensional gluons, sinusoidal lines represent the adjoint scalar $A_0$, dashed lines represent scalars, and dotted lines (not appearing in this particular figure) represent the three-dimensional ghost field.}
\label{fig:A0nSEeft}
\end{figure}

In the full theory, the (chromo)electric screening mass $m_{\rm el}^2$ is given by the position of the pole of the
propagator for the timelike component of the gauge field, $A_0^a(\tau,{\bf x})$ at spacelike momentum 
$P=(0,{\bf p})$, i.e., it is the solution to the equation
\bqa
p^2+\Pi(p^2)&=&0\;,\hspace{1cm}p^2=-m_{\rm el}^2\;,
\label{mel1}
\eqa
where $p=|{\bf p}|$, $\Pi(p^2)=\Pi_{00}(p_0=0,{\bf p})$ 
and $\Pi_{00}^{ab}(p_0,{\bf p})=\delta^{ab}\Pi_{00}(p_0,{\bf p})$
is the self-energy of the gluon field.
In ESYM, the (chromo)electric screening mass $m_{\rm el}^2$ is also given by the position of the pole of
the propagator for the adjoint field $A_0^a({\bf x})$
\bqa
p^2+m_E^2+\Pi_{\rm eff}(p^2)&=&0\;,\hspace{1cm}p^2=-m_{\rm el}^2\;,
\label{mel2}
\eqa
where $\Pi_{\rm eff}(p^2)$ is the self-energy of the adjoint scalar
in the effective theory.
By equating the expressions for the screening mass obtained by solving Eqs.~(\ref{mel1}) and~(\ref{mel2}),
we can determine the mass parameter $m_E^2$. The self-energy function 
in the full theory can be expanded in loops 
and also in a powers series around $p^2 = 0$. To leading order in the loop expansion, it suffices
to evaluate the self-energy function at $p^2=0$. In the full theory, 
the solution to Eq.~(\ref{mel1}) is
$m^2_{\rm el}=\Pi_1(0)$, where $\Pi_n(p^2)$ is of $n$th order in the loop expansion of $\Pi(p^2)$.
In the effective theory, the self-energy function evaluated at zero
external momentum vanishes in strict perturbation theory
and dimensional regularization since we are using massless propagators
and there is no scale in the loop integrals. These diagrams are shown in Fig.~\ref{fig:A0nSEeft} for completeness.

Eq.~(\ref{mel2}) then leads to $m_{\rm el}^2=m_E^2$
and therefore the mass parameter satisfies
\bqa
m_E^2&=&\Pi_1(0)\;.
\eqa

\begin{figure}[htb]
\begin{center}
\includegraphics[width=0.9\linewidth]{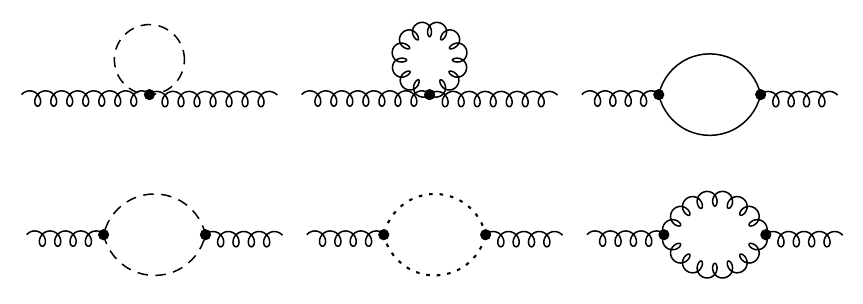}
\end{center}
\vspace{-5mm}
\caption{One-loop gluon self-energy graphs in the full $\symff$ theory.  Spiral lines represent four-dimensional gluon fields, solid lines represent four-dimensional Majorana fermions, dashed lines represent scalars, and dotted lines represent four-dimensional ghost fields.}
\label{fig:gluonSE}
\end{figure}

The $\symff$ graphs contributing to the one-loop self-energy of the zeroth component of the gauge field are shown
in Fig.~\ref{fig:gluonSE}.
It contains two parts~\cite{resumsuper}
\bqa
\Pi_{00}^{ab}(P)&=&\Pi_{00}^{{\rm b},ab}(P)+\Pi_{00}^{{\rm f},ab}(P)\;,
\eqa
where
\bqa\nonumber
\Pi_{00}^{{\rm b},ab}(P)&=&\lambda\delta^{ab}\Bigg\{
4 \sumint_Q\left[{2\over Q^2}-{(2Q_0+P_0)^2\over Q^2(P+Q)^2}
\right]
 \\ &&
-2{\bf p}^2\sumint_Q{1\over Q^2(P+Q)^2}
\Bigg\}\;,
\\  \nonumber
\Pi_{00}^{{\rm f},ab}(P)
&=&-4\lambda\delta^{ab}\Bigg\{
\sumint_{\{Q\}}\left[{2\over Q^2}-{(2Q_0+P_0)^2\over Q^2(P+Q)^2}\right]
\\ &&
- {\bf p}^2 \sumint_{\{Q\}}{1\over Q^2(P+Q)^2}\Bigg\}\;.
\eqa
After integration by parts, this yields
\bqa
m_E^2&=&\Pi_1(0)=8\lambda(d-2)(b_1-f_1)
\;.
\label{mpi1}
\eqa
The mass parameter squared $m_S^2$ can be determined along the same lines.  We define $\Sigma(p^2)$, where the self-energy of the scalar field is
$\Sigma_{AB}^{ab}(p_0=0,{\bf p})=\delta^{ab}\delta_{AB}\Sigma(p^2)$.
To leading order in the loop expansion,
we find
\bqa
m_S^2&=&\Sigma_1(0)\;,
\eqa
where $\Sigma_n(p^2)$ is of $n$th order in the loop expansion of $\Sigma(p^2)$.

\begin{figure}[htb]
\begin{center}
\includegraphics[width=0.8\linewidth]{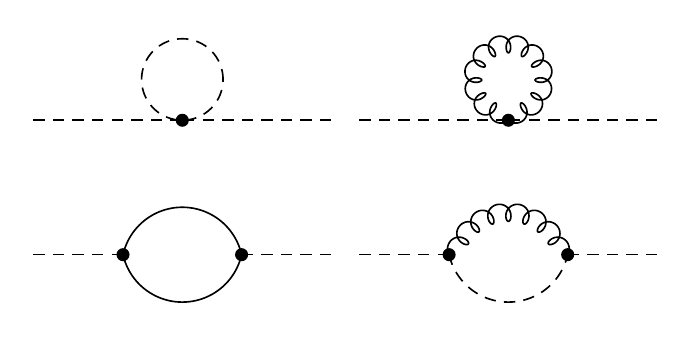}
\end{center}
\vspace{-5mm}
\caption{One-loop scalar self-energy graphs in the full $\symff$ theory. The notation is the same as in Fig.~\ref{fig:gluonSE}. }
\label{fig:scalarSE}
\end{figure}

\begin{figure}[htb]
\begin{center}
\includegraphics[width=0.75\linewidth]{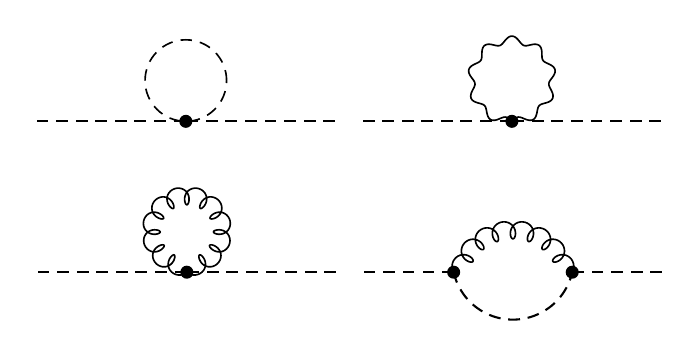}
\end{center}
\vspace{-5mm}
\caption{One-loop scalar self-energy graphs in the $\symff$ dimensionally-reduced EFT.  The notation is the same as in Fig.~\ref{fig:A0nSEeft}.}
\label{fig:scalarSEeft}
\end{figure}

The one-loop scalar self-energy graphs in the full theory are shown in Fig.~\ref{fig:scalarSE}
and their expression is~\cite{resumsuper}
\bqa\nonumber
\Sigma^{ab}_{AB}(P)&=&\lambda\delta^{ab}\delta_{AB}
\left\{2\sumint_Q\left[{4\over Q^2}-{P^2\over Q^2(P+Q)^2}\right]
\right. \\ &&\left.
-4\sumint_{\{Q\}}\left[{2\over Q^2}-{P^2\over Q^2(P+Q)^2}\right]
\right\}\;.
\eqa
For completeness, we also show the corresponding graphs in the effective theory
in Fig.~\ref{fig:scalarSEeft}, although the diagrams vanish identically in strict perturbation when evaluated at zero external momentum.
The matching then yields
\bqa
m_S^2&=&
\label{ms1}
8\lambda(b_1-f_1)\;.
\eqa
The mass parameters $m_E$ and $m_S$ are independent of the renormalization scale $\Lambda$ to this order
in strict perturbation theory.

\subsection{Coupling constants}

In order to calculate the free energy through order $\lambda^2$, the couplings 
$g_E$, $g_3$, and $h_E$ are needed at tree level only. 
The matching is fairly straightforward since we can read off the
couplings from the full theory. We simply make the substitution 
$A_0^a({\bf x},\tau)\rightarrow \sqrt{T}A_0^a({\bf x})$
in the full theory and compare $\int_0^{\beta} d\tau {\cal L}_{\rm SUSY}$ with
the effective theory, ${\cal L}_{\rm ESYM}$.
Tree-level matching for the gauge coupling then yields
\bqa
g_E^2&=&g^2T\;.
\label{g1}
\eqa
Proceeding in the same way, making the substitution
$\Phi_A^a({\bf x},\tau)\rightarrow\sqrt{T}\Phi_A^a({\bf x})$ and comparing the full and effective theory, we find
\bqa
g^2_3&=&g^2T\;.
\label{g2}
\eqa
Finally, we obtain
\bqa
h_E&=&g^2T\;.
\label{g3}
\eqa
The couplings $g_E$, $g_3$, and $h_E$ are all independent of the renormalization scale
$\Lambda$ to this order in strict perturbation theory.

\section{Calculations in the effective theory}
\label{sec:eft}

We have now calculated the parameters in the effective theory to the necessary order to calculate the
free energy density to order $\lambda^2$. The hard part is given above, $Tf_E$, while the
soft part is given by a two-loop calculation in the effective theory. Denoting the contribution from 
the soft scale $\sqrt{\lambda}T$ by $f_M$, we have $f_M=-{\log{\cal Z}_{\rm ESYM}\over V}$.
We have explicitly checked that the one- and two-loop contributions are independent of the
parameter $\xi$ in the class of covariant gauges.

\begin{figure}[tbh]
\begin{center}
\vspace{8mm}
\includegraphics[width=0.75\linewidth]{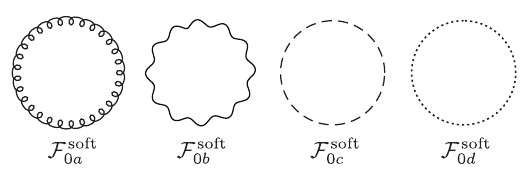}
\end{center}
\vspace{-4mm}
\caption{One-loop soft contributions to the  $\symff$ free energy density in the dimensionally-reduced EFT.  
The notation is the same as in Fig.~\ref{fig:A0nSEeft}.}
\label{fig:diagrams1eft}
\end{figure}

The one-loop graphs contributing to the free energy are shown in Fig.~\ref{fig:diagrams1eft}.
Evaluating the diagrams, we obtain
\bqa\nonumber
f_{M,1}&=&
-{1\over2}d_A\left[I_0^{\prime}(m_E^2)+6I_0^{\prime}(m_S^2)\right]
\\
&=&-{d_A\over12\pi}\left[m_E^3+6m_S^3\right] ,
\label{f1mag}
\eqa
where the integral $I_0^\prime$ is defined in App.~\ref{app:3dints} and the index after the subscript $M$ indicates the order in the loop expansion.
We note that the ghost and gauge fields are massless, which leads to a vanishing soft contribution, $I_0^{\prime}(0)=0$. 

\begin{figure}[tbh]
\begin{center}
\includegraphics[width=0.85\linewidth]{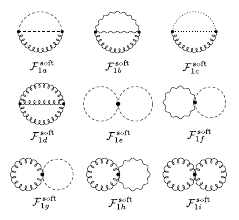}
\end{center}
\vspace{-5mm}
\caption{Two-loop soft contributions to the  $\symff$ free energy using the dimensionally-reduced EFT.  The notation is the same as in Fig.~\ref{fig:A0nSEeft}.}
\label{fig:diagrams2eft}
\end{figure}

The two-loop graphs contributing to the free energy are shown in Fig.~\ref{fig:diagrams2eft}.  Evaluating the diagrams  
and using the expressions for the integrals in Appendix B, we obtain
\begin{widetext}
\bqa\nonumber
f_{M,2}&=&d_Ag_E^2N_c\left[{1\over4}I_1^2(m_E^2)+m_E^2J_1(m_E^2)\right]+
6d_Ag_E^2N_c\left[{1\over4}I_1^2(m_S^2)+m_S^2J_1(m_S^2)\right]
\\ \nonumber && 
+3 d_Ah_EN_cI_1(m_E^2)I_1(m_S^2)
+{15\over2}d_Ag^2_3N_cI_1^2(m_S^2)\\ \nonumber
&=&
{d_AN_c\over4(4\pi)^2}\left[
{1\over\epsilon}+{3}+4\log{\Lambda\over2m_E}
\right]g_E^2m_E^2
+{3d_AN_c\over2(4\pi)^2}\left[
{1\over\epsilon}+{3}+4\log{\Lambda\over2m_S}
\right]g_E^2m_S^2
+{3d_AN_c\over(4\pi)^2}h_Em_{E}m_{S}
\\ &&
+{15d_AN_c\over2(4\pi)^2}g_3^2 m_S^2
\;,
\label{f2mag}
\eqa
\end{widetext}
Note that the setting-sun diagram with two ghost lines and one gluon line
or three gluon lines vanishes in dimensional regularization since all
the propagators are massless (diagrams ${\cal F}^{\rm soft}_{1c}$ and ${\cal F}^{\rm soft}_{1d}$). 
The same remark applies to the double bubble graphs with one or two
gluon lines (diagrams ${\cal F}^{\rm soft}_{1g}$, ${\cal F}^{\rm soft}_{1h}$, and ${\cal F}^{\rm soft}_{1i}$).
The integral $J_1(m^2)$ is logarithmically ultraviolet divergent and has a pole in $\epsilon$.
The term $f_{M,2}$ therefore requires renormalization, cf. renormalization of $f_E$.
The divergence is cancelled by the counterterm
\beq
\delta f_E = -{d_AN_c\over4(4\pi)^2\epsilon}g_E^2\left[m_E^2
+6m_S^2\right].
\label{dff}
\eeq
Comparing minimal subtraction in the full theory, Eq.~(\ref{countf}), with minimal subtraction in the effective theory, Eq.~(\ref{dff}), we see that they are not equivalent as the former
contains logarithms of the factorization scale $\Lambda$ in addition to the pole in $\epsilon$.
We note in passing that the first term in Eq.~(\ref{dff}) is the same as in QCD~\cite{braatenqcd}.
Adding Eqs.~(\ref{f2mag}) and~(\ref{dff}) yields
\begin{widetext}
\beq
f_{M,2} + \delta f_E = {d_AN_c\over(4\pi)^2}\left[
{3\over4}+\log{\Lambda\over2m_E}
\right]g_E^2m_E^2
+{d_AN_c\over(4\pi)^2}\left[
{9\over2}+6\log{\Lambda\over2m_S}
\right]g_E^2m_S^2
+{3d_AN_c\over(4\pi)^2}h_Em_{E}m_{S}
+{15d_AN_c\over2(4\pi)^2}g^2_3m_S^2\;.
\label{f22mag}
\eeq
The final result for the soft part is the sum of Eqs.~(\ref{f1mag}) and~(\ref{f22mag}). After using 
$g_E^2=g^2T$, $h_E=g^2T$, $g^2_3=g^2T$, $\lambda=g^2N_c$,
$m_E^2=2\lambda T^2$, and $m_S^2=\lambda T^2$, we find 
\bqa
f_M&=&
\label{rensoft}
-d_A{\pi^2T^3\over6}\left\{
(3+\sqrt{2})\left({\lambda\over\pi^2}\right)^{3\over2}
+ \left[
-3\log{\Lambda\over4\pi T}-{81\over16}-{9\sqrt{2}\over8}-{21\over8}\log2+{3\over2}\log{\lambda\over\pi^2}
\right]\left({\lambda\over\pi^2}\right)^{2}
\right\}\;.
\eqa
We note that the soft part Eq.~(\ref{rensoft}) explicitly depends on
the factorization scale $\Lambda$.
Adding Eqs.~(\ref{renhard}) and~(\ref{rensoft}), we obtain our final result
\bqa\nonumber
{\cal F}_{0+1+2}&=&
(f_E+f_M)T
\\ \nonumber
&=&-d_A{\pi^2T^4\over6}\left\{
1-{3\over2}{\lambda\over\pi^2}+(3+\sqrt{2})\left({\lambda\over\pi^2}\right)^{3\over2}
\right. \\ && \left.
\hspace{2cm}
+\left[-{21\over8}-{9\sqrt{2}\over8}+{3\over2}\gamma_E
+{3\over2}{\zeta^{\prime}(-1)\over\zeta(-1)}
-{25\over8}\log2+{3\over2}
\log{\lambda\over\pi^2}
\right]\left({\lambda\over\pi^2}\right)^{2}
\right\}\;.
\label{flambda2}
\eqa 
\end{widetext}
This is the complete result for the free energy through order $\lambda^2$ for general $N_c$.
It is in agreement with that of Ref.~\cite{resumsuper}, except for the finite term
$-{21\over8}$ that appears at ${\cal O}(\lambda^2)$. The reason for the difference is that one must take $d=4-2\epsilon$ in the expression for 
$\delta f_E$ in Eq.~(\ref{countf})
and not $d=4$, which gives $-{45\over16}$, as obtained in Ref.~\cite{resumsuper}.
The logarithms of the scale $\Lambda$ from the hard part 
cancel against those coming from the soft part. The absence of these logarithms
in the final result reflects that no
renormalization of the coupling is needed in SUSY. Note also the
presence of the non-analytic terms $\lambda^{3\over2}$
and $\lambda^2\log\lambda$ in Eq.~(\ref{flambda2}). 
These terms correspond to the resummation of a class
of diagrams from all orders of perturbation theory. 
The free energy density is given by Eq.~(\ref{flambda2}). All other thermodynamic quantities
can be derived from the partition function ${\cal Z}$. For example, the entropy density is
${\cal S}=-d{\cal F}/dT$.
Since the coupling $\lambda$ does not run due to conformality of SUSY, this implies that
the thermodynamic functions are of the same form when normalized to their Stefan-Boltzmann values,
e.g. ${\cal F}/{\cal F}_{\rm ideal}={\cal S}/{\cal S}_{\rm ideal}$. 

\begin{figure}[t]
\begin{center}
\includegraphics[width=0.95\linewidth]{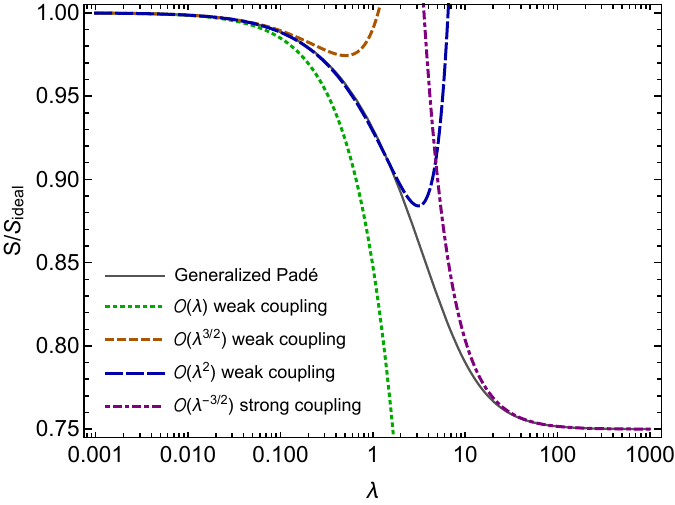}
\end{center}
\vspace{-4mm}
\caption{The entropy density ${\cal S}$ normalized by the ${\cal S}_{\rm ideal}$
in $\symff$ as a function of the 't Hooft coupling $\lambda$.
The green dotted, brown dashed, and blue long-dashed curves are the perturbative result
through order $\lambda$, $\lambda^{3\over2}$, and $\lambda^2$, respectively.
The solid grey line is the generalized Pad\'{e} approximant \eqref{eq:pade} and the purple dot-dashed curve is the strong coupling result through order $\lambda^{-{3\over2}}$.
}
\label{result}
\end{figure}

In Fig.~\ref{result}, we show the scaled entropy density as a function of $\lambda=g^2N_c$
for different truncations of the weak-coupling expansion. The green dotted, brown dashed, and
blue long-dashed curves correspond to expansions through ${\cal O}(\lambda)$, ${\cal O}(\lambda^{3\over2})$, and ${\cal O}(\lambda^2)$, respectively.  The purple dot-dashed line corresponds to the large-$N_c$ strong-coupling expansion through ${\cal O}(\lambda^{-3/2})$ and the solid grey line corresponds to a generalized [5,5] Pad\'{e} approximant which interpolates between the weak and strong coupling limits.  The analytic expression for this Pad\'{e} approximant is presented in App.~\ref{app:pade}.

In terms of convergence, our conclusions are similar to those of Ref.~\cite{resumsuper}, namely that the resummed perturbative expansion seems to be converging quickly as one adds additional perturbative orders.  One can take the value of the 't Hooft coupling at which the truncated
perturbative solutions cease to be close to the Pad\'{e} approximant as an estimate of
the range of validity of each perturbative truncation. We find that the expansions truncated at ${\cal O}(\lambda)$, ${\cal O}(\lambda^{3/2})$, and ${\cal O}(\lambda^2)$ agree well with the generalized Pad\'{e} for $\lambda \lesssim \{0.02, 0.2, 2\}$, respectively.  This suggests that the resummed perturbative expansion for $\symff$ thermodynamics has a finite and perhaps large radius of convergence.

\section{Summary and outlook}
\label{sec:conclusions}

In conclusion, we have rederived the free energy density for ${\cal N}=4$ supersymmetric Yang-Mills
theory through order $\lambda^2$ 
using dimensional reduction and effective field theory, correcting a small mistake
in the literature in the process. The weak-coupling expansion seems to have good convergence 
properties.

Dimensional reduction and effective field methods were used to streamline the calculations by explicitly separating the scales $T$ and $\sqrt{\lambda} T$.
It also explains the appearance of logarithms of the coupling $\lambda$ in the expression for the
free energy Eq.~(\ref{flambda2}). It is associated with the renormalization of the
parameter $f_E$ in the effective theory.
The solution to the renormalization group equation for these parameters can generally be used to sum leading logarithms of the form $g^{m+2n}\log^n(g)$, where $g$ again is a generic coupling~\cite{braatenscalar}. The fact that the solution to the evolution equation for 
$f_E$ is trivial suggests, as in QCD, that there are no higher order logs of the form
$\lambda^{n+1}\log^{n}\lambda$ with $n>1$
associated with the terms $\lambda^2\log\lambda$~\cite{braatenqcd}.

The next term in the weak-coupling expansion will be of order
$\lambda^{5\over2}$ and is the highest order that we can obtain in purely perturbative calculations due
to the magnetic mass problem of nonabelian gauge theories at finite temperature~\cite{linde,gross}.
The order-$\lambda^{5\over2}$ contribution to the free energy density is coming entirely from the
soft scale $\sqrt{\lambda} T$ and requires the evaluation of the three-loop vacuum diagrams in ESYM.
It also requires the determination of the mass parameters squared $m_E^2$ and $m_S^2$ to two-loop order, in analogy with the calculations in QCD~\cite{braatenqcd}.
This work in is progress~\cite{theteam}.  
Once this is complete, it would also be interesting to extend the two-loop hard-thermal-loop perturbation theory calculation of $\symff$ thermodynamics \cite{Du:2020odw} to three-loop order.

\section*{Acknowledgements}

M.S. and U.T. were supported by the U.S. Department of Energy under Award No.~DE-SC0013470. Q.D. was supported by Guangdong Major Project of Basic and Applied Basic Research No. 2020B0301030008, Natural Science Foundation of China with Project Nos. 11935007.   

\appendix

\section{Sum-integrals}
\label{app:sumints}

Loop integrals in the full theory involve sums over Matsubara frequencies and 
integrals over spatial momenta. We use momentum-space dimensional regularization to regulate both infrared
and ultraviolet divergences.
The sum-integrals are defined as
\bqa
\sumint_P&=&\left({e^{\gamma_E}\Lambda^2\over4\pi}\right)^{\epsilon}T
\sum_{p_0=2n\pi T}\int_p\;,\\
\sumint_{\{P\}}&=&\left({e^{\gamma_E}\Lambda^2\over4\pi}\right)^{\epsilon}T
\sum_{p_0=(2n+1)\pi T}\int_p\;,
\eqa
where the sum is over Matsubara frequencies, $p_0=2n\pi T$ for bosons and
$p_0=(2n+1)\pi T$ for fermions.
The integrals over momenta are denoted by
\bqa
\int_p&=&\left({e^{\gamma_E}\Lambda^2\over4\pi}\right)^{\epsilon}\int{d^{d-1}
p\over(2\pi)^{d-1}}\;,
\eqa
where $d=4-2\epsilon$ and
$\Lambda$ is an arbitrary momentum scale that coincides with the 
renormalization scale in the $\overline{\rm MS}$ scheme.

The simple one-loop sum-integrals are of the form
\bqa
b_n&=&\sumint_P{1\over P^{2n}}
\;,
\hspace{0.4cm}f_n=\sumint_{\{P\}}{1\over P^{2n}}\;,
\;\;\;
n\geq0\;.
\label{bfdef}
\eqa
We specifically need the following one-loop sum-integrals
\bqa
\label{b0}
b_0^{\prime}&=&{\pi^2\over45}T^4\left[1+{\cal O}(\epsilon)\right],\\
\label{b1}
b_1&=&{T^2\over12}\left({\Lambda\over4\pi T}\right)^{2\epsilon}
\left[
1+\left(2+2{\zeta^{\prime}(-1)\over\zeta(-1)}\right)\epsilon+{\cal O}(\epsilon^2)
\right] , \nonumber \\ && \\
b_2&=&
{1\over(4\pi)^2}\left({\Lambda\over4\pi T}\right)^{2\epsilon}\left[{1\over\epsilon}
+2\gamma+{\cal O}(\epsilon)\right] ,
\label{b2}
\eqa

\bqa
\label{f0}
f_0^{\prime}&=&-
{7\pi^2\over360}T^4\left[1+{\cal O}(\epsilon)\right],\\ \nonumber
f_1&=&-{T^2\over24}\left({\Lambda\over4\pi T}\right)^{2\epsilon}
\bigg[
1+\left(2-2\log2+2{\zeta^{\prime}(-1)\over\zeta(-1)}\right)\epsilon
\\ &&
\hspace{3cm} +{\cal O}(\epsilon^2)
\bigg]\;,\label{f1} \\
f_2&=&
{1\over(4\pi)^2}\left({\Lambda\over4\pi T}\right)^{2\epsilon}\left[{1\over\epsilon}
+4\log2+2\gamma+{\cal O}(\epsilon)\right] ,
\label{f2}
\eqa
where the prime indicates a derivative with respect to the exponent $n$ in Eq.~(\ref{bfdef}).

The following two-loop sum-integrals arise in the simplification of certain three-loop diagrams in the full theory and vanish~\cite{arnold1}
\bqa
\sumint_{PQ}{1\over P^2Q^2(P+Q)^2}&=&0\;,\\
\sumint_{\{P\}Q}{1\over P^2Q^2(P+Q)^2}&=&0\;,\\
\sumint_{\{PQ\}}{1\over P^2Q^2(P+Q)^2}&=&0\;.
\eqa
\begin{widetext}

The three-loop sum-integrals needed are
\bqa
\label{bb}
I_{\rm ball}^{\rm bb}&=&
{1\over(4\pi)^2}\left({T^2\over12}\right)^2\left[{6\over\epsilon}+36\log{\mu\over4\pi T}
-12{\zeta^{\prime}(-3)\over\zeta(-3)}+48{\zeta^{\prime}(-1)\over\zeta(-1)}+{182\over5}
+{{\cal O}(\epsilon)}\right] ,\\
\label{ff}
I_{\rm ball}^{\rm ff}&=&
{1\over(4\pi)^2}\left({T^2\over12}\right)^2\left[{3\over2\epsilon}+9\log{\mu\over4\pi T}
-3{\zeta^{\prime}(-3)\over\zeta(-3)}+12{\zeta^{\prime}(-1)\over\zeta(-1)}
+{173\over20}-{63\over5}\log2
+{{\cal O}(\epsilon)}\right] ,\\
\label{bf}
I_{\rm ball}^{\rm bf}&=&-{1\over6}(1-2^{11-3d})I_{\rm ball}^ {\rm bb}
-{1\over6}I_{\rm ball}^ {\rm ff}
\;,\\ 
H_3&=&
{1\over(4\pi)^2}\left({T^2\over12}\right)^2
\left[{3\over8\epsilon}+{9\over4}\log{\mu\over4\pi T}
+{3\over2}{\zeta^{\prime}(-3)\over\zeta(-3)}-{3\over2}{\zeta^{\prime}(-1)\over\zeta(-1)}
+{9\over4}\gamma_E
+{361\over160}+{57\over10}\log2+{\cal O}(\epsilon)\right],
\label{h3}
\\
H_4&=&
{1\over(4\pi)^2}\left({T^2\over12}\right)^2
\left[{5\over24\epsilon}+{5\over4}\log{\mu\over4\pi T}
-{1\over6}{\zeta^{\prime}(-3)\over\zeta(-3)}+{7\over6}{\zeta^{\prime}(-1)\over\zeta(-1)}+{1\over4}\gamma_E
+{23\over24}-{8\over5}\log2+{\cal O}(\epsilon)\right],
\label{h4}
\\
H_5&=&
{1\over(4\pi)^2}\left({T^2\over12}\right)^2\left[{4\over3\epsilon}+8\log{\mu\over4\pi T}
-{5\over3}{\zeta^{\prime}(-3)\over\zeta(-3)}+{26\over3}{\zeta^{\prime}(-1)\over\zeta(-1)}+\gamma_E
+{49\over12}+{{\cal O}(\epsilon)}\right],\\ 
\label{h5}
H_6&=&
{1\over(4\pi)^2}\left({T^2\over12}\right)^2\left[-{17\over48\epsilon}-{17\over8}\log{\mu\over4\pi T}
+{5\over24}{\zeta^{\prime}(-3)\over\zeta(-3)}-{11\over6}{\zeta^{\prime}(-1)\over\zeta(-1)}
-{1\over2}\gamma_E
-{41\over48}+{11\over8}\log2+{{\cal O}(\epsilon)}\right] ,
\label{h6}
\eqa
\end{widetext}
and
\bqa
I_{\symod}^{\textrm{bb}} &=& \frac{({\mathcal D}-2)^2}{4} \bar{I}^{\textrm{bb}}_{\symod} + 2 {\mathcal D} I_{\textrm{ball}}^{\textrm{bb}} \, , \\
I_{\symod}^{\textrm{ff}} &=& \frac{{({\rm Tr}\,\mathbb{1})^2}}{4} \big[ \bar{I}^{\textrm{ff}}_{\symod} +({\mathcal D}-3)I_{\textrm{ball}}^{\textrm{ff}} \big] ,\\
I_{\symod}^{\textrm{bf}} &=& -{\textrm{Tr} \, \mathbb{1}} \bigg[ \frac{{\mathcal D} -2}{2}\bar{I}^{\textrm{bf}}_{\symod} + \frac{3}{2}({\mathcal D}-2)I_{\textrm{ball}}^{\textrm{bf}}  \bigg] , \nonumber \\
\eqa
with
\bqa
\bar{I}^{\textrm{bb}}_{\symod} &=& 4({\mathcal D}-4)b_2 b_1^2 +16 H_5-I_{\textrm{ball}}^{\textrm{bb}} \, ,\\
\bar{I}^{\textrm{ff}}_{\symod} &=& 4({\mathcal D}-4)b_2 f_1^2 +16 H_4-I_{\textrm{ball}}^{\textrm{ff}}  \, ,\\
\bar{I}^{\textrm{bf}}_{\symod} &=&  4({\mathcal D}-4)b_2 b_1 f_1 +16 H_6 - I_{\textrm{ball}}^{\textrm{bf}} \, .
\eqa
The three-loop sum-integrals in Eqs.~(\ref{bb})--(\ref{h3}) were calculated in 
Refs.~\cite{arnold1,arnold2}. The remaining three-loop sum-integrals were calculated in Ref.~\cite{resumsuper}.

\section{Integrals in the effective theory}
\label{app:3dints}

Loop diagrams in the effective three-dimensional theory involve integrals over
three-momenta. We use dimensional regularization to regulate both infrared and 
ultraviolet divergences. The integrals are denoted by
\bqa
\int_p&=&\left({e^{\gamma_E}\Lambda^2\over4\pi}\right)^{\epsilon}
\int {d^{d-1} p \over(2\pi)^{d-1}}
\;,
\eqa
where $d=4-2\epsilon$  and $\Lambda$
is the renormalization scale in 
the $\overline{\rm MS}$ scheme. 
We define the one-loop integrals
\bqa
I_n(m^2)&=&\int_p{1\over[p^2+m^2]^n}\;.
\label{in}
\eqa
The specific one-loop integrals we need are
\bqa
I_0^{\prime}(m^2)&=&{m^3\over4\pi}\left({\Lambda\over2m}\right)^{2\epsilon}\left[{2\over3}
+{16\over9}\epsilon+{\cal O}(\epsilon^2)\right],
\\
I_1(m^2)&=&{m\over4\pi}\left({\Lambda\over2m}\right)^{2\epsilon}\left[-1
-2\epsilon+{\cal O}(\epsilon^2)\right]\;,
\eqa
where the prime again indicates the derivative with respect to the exponent $n$ in Eq.~(\ref{in}).

Some of the two-loop diagrams are simple products of the one-loop integrals
defined in Eq.~(\ref{in}). 
The two-loop integrals that are not simple products are of the form~\cite{braatenqcd}
\bqa
J_n(m^2)&=&\int_{pq}{1\over(p^2+m^2)[q^2+m^2]^n(p-q)^2}\;.
\eqa
Specifically, we need the two-loop diagram
\bqa
J_1(m^2)&=&
{1\over(4\pi)^2}\left({\Lambda\over2m}\right)^{4\epsilon}\left[{1\over4\epsilon}
+{1\over2}+{\cal O}(\epsilon)\right].
\eqa

\begin{widetext}

\section{Generalized Pad\'{e}}
\label{app:pade}

Following Ref.~\cite{resumsuper} one can construct a generalized Pad\'{e} approximant that interpolates between the known weak- and strong-coupling limits.  We find that in the large $N_c$-limit the following form 
\beq
\frac{S}{S_{\rm ideal} } =  \frac{1 + a \lambda^{1/2}  + b \lambda  + c \lambda^{3/2}  + d \lambda^2  + e \lambda^{5/2}  }{1 + a \lambda^{1/2}  + \bar{b} \lambda  + \frac{4}{3} c \lambda^{3/2}  + \frac{4}{3} d \lambda^2  + \frac{4}{3} e \lambda^{5/2}  }  \; , 
\label{eq:pade}
\eeq
with
\bqa
a &=& \frac{4 \pi ^2}{135 \zeta (3)}+\frac{2 \left(3+\sqrt{2}\right)}{3 \pi } \, , \nonumber \\
b &=& 
\frac{1}{\pi^2} \log\left( \frac{\lambda}{\pi^2} \right) +\frac{16 \pi  \left[ 45 \left(3+\sqrt{2}\right) \zeta (3)+\pi
   ^3\right]}{18225 \zeta^2(3)}
   +\frac{36 \left[ \frac{\zeta'(-1)}{\zeta(-1)}+ \gamma \right] +69 \sqrt{2}+59-75 \log 2}{36 \pi ^2} \, , \nonumber  \\
&& \nonumber  \\
\bar{b} &=& b + \frac{3}{2 \pi ^2} \, ,  \nonumber  \\
c &=& \frac{2}{15 \zeta (3)} \, , \nonumber  \\
d &=& \frac{180 \left(3+\sqrt{2}\right) \zeta (3)+8 \pi ^3}{2025 \pi  \zeta^2(3)} \, , \nonumber  \\
e &=& \frac{2b}{15 \zeta (3)}-\frac{3}{5 \pi ^2 \zeta (3)} \, ,
\eqa
reproduces Eqs.~\eqref{eq:sclimit} and \eqref{flambda2} in the strong- and weak-coupling limits, respectively, and that all coefficients are uniquely constrained.  For details concerning the method of construction see Ref.~\cite{resumsuper}. Note that this is different than the result originally reported in Ref.~\cite{resumsuper} in the last term contributing to the coefficient $b$.  This has been corrected in an erratum to Ref.~\cite{resumsuper}.

\end{widetext}

\end{document}